\documentclass[11pt,english,pre,preprint,superscriptaddress,floatfix]{revtex4}
\usepackage[T1]{fontenc}
\usepackage[latin9]{inputenc}
\usepackage{color}
\usepackage{babel}
\usepackage{units}
\usepackage{amsmath}
\usepackage{graphicx}
\usepackage{esint}
\PassOptionsToPackage{normalem}{ulem}
\usepackage{ulem}
\usepackage[unicode=true,pdfusetitle,
 bookmarks=true,bookmarksnumbered=false,bookmarksopen=false,
 breaklinks=false,pdfborder={0 0 1},backref=false,colorlinks=false]
 {hyperref}

\makeatletter
\@ifundefined{textcolor}{}
{%
 \definecolor{BLACK}{gray}{0}
 \definecolor{WHITE}{gray}{1}
 \definecolor{RED}{rgb}{1,0,0}
 \definecolor{GREEN}{rgb}{0,1,0}
 \definecolor{BLUE}{rgb}{0,0,1}
 \definecolor{CYAN}{cmyk}{1,0,0,0}
 \definecolor{MAGENTA}{cmyk}{0,1,0,0}
 \definecolor{YELLOW}{cmyk}{0,0,1,0}
}

\makeatother

\begin{document}

\title{Theory and learning protocols for the material tempotron model}

\author{Carlo Baldassi}

\affiliation{DISAT and Center for Computational Sciences, Politecnico di Torino,
Corso Duca degli Abruzzi 24, 10129 Torino, Italy}

\affiliation{Human Genetics Foundation, Via Nizza 52, 10126 Torino, Italy}

\author{Alfredo Braunstein}

\affiliation{DISAT and Center for Computational Sciences, Politecnico di Torino,
Corso Duca degli Abruzzi 24, 10129 Torino, Italy}

\affiliation{Human Genetics Foundation, Via Nizza 52, 10126 Torino, Italy}

\affiliation{Collegio Carlo Alberto, Via Real Collegio 30, 10024 Moncalieri, Italy}

\author{Riccardo Zecchina}

\affiliation{DISAT and Center for Computational Sciences, Politecnico di Torino,
Corso Duca degli Abruzzi 24, 10129 Torino, Italy}

\affiliation{Human Genetics Foundation, Via Nizza 52, 10126 Torino, Italy}

\affiliation{Collegio Carlo Alberto, Via Real Collegio 30, 10024 Moncalieri, Italy}
\begin{abstract}
Neural networks are able to extract information from the timing of
spikes. Here we provide new results on the behavior of the simplest
neuronal model which is able to decode information embedded in temporal
spike patterns, the so called tempotron \cite{gutig_tempotron:_2006}.
Using statistical physics techniques we compute the capacity for the
case of sparse, time-discretized input, and ``material'' discrete
synapses, showing that the device saturates the information theoretic
bounds with a statistics of output spikes that is consistent with the
statistics of the inputs. We also derive two simple and highly
efficient learning algorithms which are able to learn a number of
associations which are close to the theoretical limit. The simplest
versions of these algorithms correspond to distributed on-line
protocols of interest for neuromorphic devices, and can be adapted to
address the more biologically relevant continuous-time version of the
classification problem, hopefully allowing for the understanding of
some aspects of synaptic plasticity.  \end{abstract} \maketitle
\tableofcontents{}

\section{Introduction\label{sec:Introduction}}

Recent studies concerning learning processes in neural circuits have
highlighted the role of spike timing and synchrony (e.g.~in sensory
systems \cite{johansson_first_2004,decharms_primary_1996,meister_concerted_1995,wehr_odour_1996}),
leading to a view of the learning devices as a class of time coincidence
detectors of a limited number of spikes (at least under certain circumstances).
These observations are at the root of several fundamental questions
concerning neural coding, the most important one possibly being how
do neurons learn to recognize multiple spatiotemporal patterns.

A very stimulating contribution in this field has been the recent
introduction by Gutig and Sompolinsky \cite{gutig_tempotron:_2006}
of a perceptron-like model neuron which is able to process spatio-temporal
patterns, the so called tempotron. In spite of its simplicity, such
a device is capable of decoding the information contained in the synchrony
of spike patterns through a relatively simple supervised gradient
learning rule. Subsequent work \cite{rubin_theory_2010} has analyzed
by statistical physics techniques the storage capacity (i.e.~the
typical maximum number of distinct input-output associations which
the device can in principle be trained to reproduce, assuming that
the inputs and the expected responses are drawn from some probability
distribution) and the geometry of the space of solutions of the tempotron
for continuous synaptic weights. 

In a nutshell, the tempotron is the simplest form of an integrate
and fire (IF) neuron, with $N$ input synapses of strength $J_{i}$,
$i=1,\dots,N$ (also called synaptic weights).
In the tempotron, each input corresponds to $N$ sequences
of spikes, where the set of spiking times is denoted by $\{t_{i}\}$.
The tempotron performs a binary classification of the inputs depending
on whether the membrane potential reaches or not the firing threshold
$\theta$ in the given time interval. The potential at time $t$ is
given by $V\left(t\right)=\sum_{i}J_{i}\sum_{t_{i}<t}v\left(t-t_{i}\right)$, where $v\left(t\right)$
is the temporal kernel of the membrane. A standard choice for the
kernel is the exponential one, namely $v\left(t\right)=v_{0}\left(e^{-t/t_{m}}-e^{-t/t_{s}}\right)$,
where $t_{m}$ and $t_{s}$ are the membrane and synaptic integration
time constants. For this model the precise timings and the number
of output spikes (if greater than one) play no role in the binary
classification, allowing for multiple equivalent output spiking
profiles for positive classifications of a given time interval.

As discussed in \cite{rubin_theory_2010}, a key parameter is the
quantity $K=T/\sqrt{t_{m}t_{s}}$ where $T$ is the duration of the
input pattern. When both $N$ and $K$ are large, and $N\gg K$,
certain time correlations can be neglected and
the analysis simplifies. This has allowed the authors of \cite{rubin_theory_2010}
to estimate the storage capacity of the device for the case of continuous
weights and random i.i.d.\ patterns. It is interesting to observe that
these conditions are not far from being actually realistic
\cite{rubin_theory_2010}. The vanishing of the correlations at different
times is due to the sparse regime under which the device operates, and
it means that the width of the kernel $v\left(t\right)$ is much shorter than
the typical interval between incoming spikes; this in turn means that,
under this regime, only quasi-simultaneous input spikes actually contribute
to the depolarization at any given time, which explains why in~\cite{rubin_theory_2010}
the theoretical and experimental results are closely reproduced
with a simplified model in which time is discretized in $K^\textrm{discrete}=K/8$
bins and the output is simply given by a perceptron rule applied on each bin
(see paragraph~\ref{par:Model} for additional details).

Basic devices like the temportron have the potential virtue of touching
those fundamental questions in neural coding which are preserved in
spite of the simplicity of the device itself. In such a framework,
we have approached the problem from a different angle, namely adopting
a computational scheme which that is not based on a gradient-like
computation but is still fully local and distributed. We study the
simplified (time-discretized) tempotron by both the replica method and
the so called message-passing approach (or cavity method) which allows
us to study analytically the storage capacity and at the same time to
derive simple learning protocols (i.e.~training rules for the
modifications of the synaptic strengths, which the device applies upon
receiving input patterns and being made aware of the expected
response, such that at the end of the training the desired set of
input-output associations is learned) which are efficient and do not
rely on any continuity condition of the synaptic weights. We also show
how to adapt the simplest of these learning protocols to address the
original, continuous-time model. For the sake of simplicity we focus
directly on the case of discrete synapses, although the results could
be extended to the continuous case.

For the cases of single and multilayer perceptrons with firing rate
coding and binary synapses we have shown in previous works \cite{braunstein_learning_2006,baldassi_efficient_2007,baldassi_generalization_2009}
that the message-passing approach is indeed efficient in solving the
learning problem for random patterns and that the computational scheme
can be simplified to the point of providing extremely simple learning
protocols. These past results together with the novel ones on spatio-temporal
coding presented here should be of practical interest for large scale
neuromorphic devices and hopefully for providing novel hints on aspects
of synaptic plasticity.

\paragraph{The model\label{par:Model}}

We studied two tempotron scenarios, one in which synaptic conductances
can take values in $\left\{ -1,+1\right\} $, and one in which they
can take values in $\left\{ 0,1\right\} $, focusing on the former
case in simulations. As in the final paragraphs of~\cite{rubin_theory_2010}, we worked under
the simplifying assumption that input and output spike patterns can
be encoded (via binning) as sparse strings of $0$'s and $1$'s, and
that the relationship between the inputs and the output at any given
time bin is given by a perceptron rule. As mentioned
in~\cite{rubin_theory_2010} and in the Introduction, we expect that
this simplification does not qualitatively alter the overall picture
under the sparse regime considered, since even in the
integrate-and-fire model only quasi-simultaneous input spikes affect
the overall depolarization at any given time, due to the fast membrane
decay constant w.r.t.\ the typical inter-spike interval; indeed, our
numerical results (see section \ref{sub:Continuous}) show that it is even
possible to use a time-discretized learning protocol to address the
original continuous-time classification problem.

We thus consider a classification
device with $N$ binary synapses, $J_{i}$ with $i\in\left\{ 1,\dots,N\right\} $,
which has to learn to classify $M=\alpha N$ input patterns. The input
patterns are $N\times K$ matrices (where $K$ here corresponds to the $K^\textrm{discrete}$
discussed in the Introduction) whose elements are $\xi_{it}^{\mu}\in\left\{ 0,1\right\} $
with $i\in\left\{ 1,\dots,N\right\} $, $t\in\left\{ 1,\dots,K\right\} $
and $\mu\in\left\{ 1,\dots,M\right\} $. For each pattern $\mu$ and
at each time step $t$, the device response is given by $V_{t}^{\mu}=\Theta\left(\sum_{i=1}^{N}J_{i}\xi_{it}^{\mu}-\theta\right)$,
where $\Theta\left(x\right)$ is the Heaviside step function and $\theta$
is a threshold; we call the vector $V^{\mu}=\left\{ V_{t}^{\mu}\right\} _{t\in\left\{ 1,\dots,K\right\} }$
the \emph{internal representation} for the pattern $\mu$. Finally,
a pattern $\mu$ is classified according to $s^{\mu}=1-\prod_{t=1}^{K}\left(1-V_{t}^{\mu}\right)$,
i.e.~the overall output $s^{\mu}$ equals $0$ if the internal representation
is a vector of all $0$'s, or it equals $1$ if at least one element
of $V^{\mu}$ is $1$. Each pattern has a desired output $s_{\textrm{exp}}^{\mu}\in\left\{ 0,1\right\} $
which is to be compared to the actual output $s^{\mu}$: the classification
problem is satisfied when $s^{\mu}=s_{\textrm{exp}}^{\mu}$ for all
$\mu$. For notational simplicity, we also define $\sigma^{\mu}=2s^{\mu}-1$
and $\sigma_{\textrm{exp}}^{\mu}=2s_{\textrm{exp}}^{\mu}-1$ when
we need to convert the outputs so that they take values in $\left\{ -1,+1\right\} $.

We studied the case in which all inputs and expected outputs are i.i.d.\ random
variables. We call $f^{\prime}$ the output frequency (i.e.~$s_{\textrm{exp}}^{\mu}=1$
with probability $f^{\prime}$) and $f$ the input frequency (i.e.~$\xi_{it}^{\mu}=1$
with probability $f$). We will assume in the following that $f=\left(1-\left(1-f^{\prime}\right)^{\frac{1}{K}}\right)$:
this ensures that the probability that a vector $\left\{ \xi_{it}^{\mu}\right\} _{t\in\left\{ 1,\dots,K\right\} }$
is composed of all $0$'s is $\left(1-f^{\prime}\right)$, i.e.~has
the same statistics as the internal representations which satisfy
the input/output associations. Clearly, the model reduces to a standard
perceptron when $K=1$.

We assume that $N\gg1$, $K\gg1$ and $N\gg K$; in this case, $f\simeq-K^{-1}\log\left(1-f^{\prime}\right)$,
which shows that the inputs are sparse at large $K$ with our choice
for $f$. For simplicity, all our theoretical results and simulations
will be presented for the case $f^{\prime}=0.5$, which is the value 
that maximizes the capacity.

\section{Theoretical analysis}

\subsection{Replica theory}

We studied the device described above within the replica theory, in
a replica symmetric (RS) setting for the internal representations,
in the limit of large $K$, both for the case in which $J_{i}\in\left\{ -1,+1\right\} $
and $J_{i}\in\left\{ 0,1\right\} $, and estimated the entropy, the
critical capacity, the optimal value for the threshold, and studied
the structure of the space of the solutions and the valid internal
representations. The results are almost identical for both $\pm1$
and $0/1$ cases, so in the following we will only specify the model
when a difference arises. We confirmed the results, where possible,
with the cavity method (see section~\ref{sub:Cavity-method}). All
details of the calculations are provided in the Appendix (section~\ref{sec:Appendix:-replica-calculations});
here we summarize the results.

The zero-temperature entropy of the device is defined as
$S\left(\alpha\right) = \frac{1}{N}\left<\log\left(\mathcal{V}\right)\right>$,
where $\mathcal{V}$ is the number of solutions (valid configurations of $J$'s) to the problem associated
with some choice of the patterns $\xi_{it}^\mu$ and their expected outputs $s_{\textrm{exp}}^\mu$ (see also
eq.~\ref{eq:V}), and $\left<\cdot\right>$ denotes the average over the patterns.
$S\left(\alpha\right)$ can't be negative (since the number of valid configurations is an integer number);
the value of $\alpha$ at which $S\left(\alpha\right)$ vanishes is called the \emph{critical capacity},
and represents the typical number of patterns per synapse which can be correctly classified by the device
(i.e.~stored) when the patterns are extracted according to the random i.i.d. distribution which
we are studying.

The RS replica calculation predicts $S\left(\alpha\right) = \log\left(2\right)\left(1-\alpha\right)$,
which interestingly does not depend on $K$ (provided $K \gg 1$). This function
goes to zero at $\alpha_c = 1$, which coincides with the information theoretic
upper bound, i.e.~the device is able to store one bit of information per synapse.
This is in contrast with other related architectures, e.g.~the multi-layer perceptron.
After this point, the entropy is negative, and therefore the RS solution is no longer valid.

The typical value of the overlap between two different solutions to
the same classification problem, defined as $q=\frac{1}{N}\left<\sum_{i=1}^{N}J_{i}^{a}J_{i}^{b}\right>$
for two solutions $\left\{ J_{i}^{a}\right\} _{i\in\left\{ 1,\dots,N\right\} }$
and $\left\{ J_{i}^{b}\right\} _{i\in\left\{ 1,\dots,N\right\} }$,
is constant for all values of $\alpha$, and as low as possible,
i.e.~$q=0$ in the $\pm1$ case and $q=Q^{2}=\nicefrac{1}{4}$
in the $0/1$ case, where $Q$ is the typical fraction of non-null
synapses (which we found to be $Q=\nicefrac{1}{2}$). In terms of
the structure of the space of the solutions, this means that the clusters
of solutions are isolated (point-like).

We expand the threshold in series of $\sqrt{N}$ and write $\theta=\theta_{0}N+\theta_{1}\sqrt{N}$.
The optimal value of $\theta_{0}$ is $0$ in the $\pm1$ case and
$\frac{f}{2}$ for the $0/1$ case; the value of $\theta_{1}$ does
not affect the capacity, but we can set it so that synaptic values
are unbiased: 
\begin{equation}
\theta_{1}=-\sqrt{2f\left(1-f\right)}\textrm{erfc}^{-1}\left(2\sqrt[K]{1-f^{\prime}}\right)\label{eq:opt_theta}
\end{equation}
where $\textrm{erfc}^{-1}$ is the inverse of the complementary error
function.

We found that the valid internal representations follow a binomial
distribution at large $K$, i.e.~that the probability distribution
for the value at each time bin is independent of the others. This fact
is in agreement with the continuous model findings
\cite{rubin_theory_2010}, and it is interesting for two reasons: on
one hand, it confirms that different time bins are uncorrelated in the
sparse limit, which is important in order to achieve efficiency in
applying the cavity method.  On the other hand, it means that the
distributions of the input and output spike trains are identical (note
that our choice of $f$ only ensures that the all-zero string occur
with the same probability in input and output, but does not imply that
the non-zero strings have the same distribution of the number of
spikes).  In turn, this is a necessary condition for recurrent
networks to be built and work under this regime, which would be an
interesting direction for future research.

We also computed how the internal representations are partitioned,
and found that the rescaled entropy of the dominant internal representations
(i.e.~the logarithm of the number of different internal representations
for a pattern which are associated with the largest portions of the
solution space) is given by $\log\left(2\right)\log\left(\frac{K}{\log\left(2\right)}\right)$.
This means that, as $K$ increases, the number of valid dominant internal
representations increases as $K^{M\,\log2}$, while the number of
synaptic states associated with each of them correspondingly shrinks,
so that the overall entropy remains constant.

\subsection{Cavity method\label{sub:Cavity-method}}

The cavity method has been shown~\cite{mezard_space_1989} to provide an alternative
scheme for deriving the results from replica theory in the case of the
binary perceptron with binary $\pm1$ inputs and binary $\pm1$ synapses.
It has also been used on single instances of the learning problem on such
devices (in which case it is known as the Belief Propagation algorithm,
i.e.~BP) to study the space of the solutions for some particular instance
and for deriving heuristic learning algorithms~\cite{braunstein_learning_2006,baldassi_efficient_2007}.
In those studies, the problem is represented as a factor graph in which
synaptic weights are represented by variable nodes, and input patterns
(and their desired outputs) are represented by factor nodes; messages are
exchanged between the two types of nodes along the edges of the graph,
representing marginal probabilities over the states of the variables;
global thermodinamic quantities such as the entropy can be computed from
the messages provided they satisfy the BP equations (which is typically
achieved by reaching a fixed point in an iterative algorithm). Replica theory
results can then be reproduced numerically with BP by averaging the computed
quantities from a large number of samples of sufficient size.

In the case of the present study, however, in which the input patterns take values in
$\left\{0,1\right\}$, the approach used in \cite{braunstein_learning_2006}
can not be applied directly for the sake of reproducing replica theory results,
not even in the perceptron limit $K=1$. This is due to a violation of the underlying
assumption of the cavity method, known as the clustering property, as will be explained
in greater detail at the end of this section, and as a result the standard BP
equations are approximate, rather than becoming asymptotically
exact in the large $N$ regime. Thus, in order to reproduce
the replica theory results, the standard BP equations must be amended;
however, since the heuristic algorithm described in section~\ref{sub:reinBP} is based on
the standard BP, which is simpler and more computationally efficient, and
proves equally effective to the corrected-BP version, we will first derive
the standard BP equations, and describe how to correct them afterwards.

\paragraph{Standard Belief Propagation algorithm\label{par:Standard-BP}}

As mentioned above, messages on the factor graph represent probabilities
over the variable nodes (i.e.~the synaptic weights), and therefore
can be represented by a single real value: as in~\cite{braunstein_learning_2006},
we use average values for this purpose (also called magnetizations).

The BP equations, written in terms of the cavity magnetizations $m$
and $n$, read:
\begin{eqnarray}
m_{i\to\mu} & = & \tanh\left(\sum_{\nu\ne\mu}\tanh^{-1}\left(n_{\nu\to i}\right)\right)\label{eq:m1}\\
n_{\mu\to i} & \propto & P\left(\sigma_{\textrm{exp}}^{\mu}=1-\prod_{t=1}^{K}\Theta\left(\theta-\sum_{j\ne i}J_{j}\xi_{jt}^{\mu}-\xi_{it}^{\mu}\right)\right)+\label{eq:n1}\\
 &  & -P\left(\sigma_{\textrm{exp}}^{\mu}=1-\prod_{t=1}^{K}\Theta\left(\theta-\sum_{j\ne i}J_{j}\xi_{jt}^{\mu}+\xi_{it}^{\mu}\right)\right)\nonumber 
\end{eqnarray}
where $i,j$ are synapse indices and $\mu,\nu$ are pattern indices.
The second equation is the difference between the probabilities that
the pattern $\mu$ is satisfied when synapse $i$ takes the values
$1$ and $-1$, respectively, assuming all other synaptic values are
distributed according to the cavity magnetizations $m_{j\to\mu}$
(for $j\ne i$). These can be computed from the probability that the
internal representation is all zero given the value of $J_{i}$:
\begin{equation}
B_{\mu\to i}\left(J_{i}\right)=\sum_{\left\{ J_{j}\right\} _{j\ne i}}\prod_{j\ne i}\left(\frac{1}{2}+J_{j}\frac{\mu}{2}\right)\prod_{t=1}^{K}\Theta\left(\theta-\sum_{j\ne i}J_{j}\xi_{jt}^{\mu}-J_{i}\xi_{it}^{\mu}\right)\label{eq:B1}
\end{equation}

With this, and using the shorthand notation $B_{\mu\to i}^{\pm}=B_{\mu\to i}\left(\pm1\right)$,
we can write eq.~\ref{eq:n1} as:
\begin{equation}
n_{\mu\to i}=\frac{B_{\mu\to i}^{+}-B_{\mu\to i}^{-}}{B_{\mu\to i}^{+}+B_{\mu\to i}^{-}}\left(1-\frac{2s_{\textrm{exp}}^{\mu}}{2-B_{\mu\to i}^{+}-B_{\mu\to i}^{-}}\right)\label{eq:n2}
\end{equation}

In order to compute efficiently the function $B$, we use the central
limit theorem, which ensures that for large $N$ we have:
\begin{equation}
B_{\mu\to i}\left(J_{i}\right)=\int_{\mathcal{S}_{\mu\to i}\left(J_{i}\right)}\prod_{t=1}^{K}dy_{t}\ \mathcal{N}\left(\bar{y};\bar{a}_{\mu\to i},\bar{\Sigma}_{\mu\to i}\right)\label{eq:B2}
\end{equation}
where $\mathcal{N}\left(\bar{y};\bar{a},\bar{\Sigma}\right)$ is a
$K$-dimensional multivariate Gaussian with mean $\bar{a}$ and covariance
matrix $\bar{\Sigma}$, whose elements are given by:
\begin{eqnarray}
\left(\bar{a}_{\mu\to i}\right)_{t} & = & \sum_{j\ne i}\xi_{jt}^{\mu}m_{j\to\mu}\label{eq:a1}\\
\left(\bar{\Sigma}_{\mu\to i}\right)_{tt^{\prime}} & = & \sum_{j\ne i}\xi_{jt}^{\mu}\xi_{jt^{\prime}}^{\mu}\left(1-m_{j\to\mu}^{2}\right)\label{eq:Sigma1}
\end{eqnarray}

The region of integration is a product of semi-bounded intervals:
$\mathcal{S}_{\mu\to i}\left(J_{i}\right)=\bigotimes_{t=1}^{K}\left(-\infty,\theta-J_{i}\xi_{it}^{\mu}\right]$.

Computing this integral in general is very expensive, and rapidly
becomes infeasible for large $K$. However, the sparsity of the input
patterns implies that diagonal terms are of order $K^{-1}$, while
off-diagonal terms are of order $K^{-2}$ and can be neglected, simplifying
the computation:
\begin{equation}
B_{\mu\to i}\left(J_{i}\right)=\prod_{t=1}^{K}\frac{1}{2}\textrm{\textrm{erfc}}\left(\frac{1}{\sqrt{2}}\left(\frac{\theta-J_{i}\xi_{it}^{\mu}-\left(\bar{a}_{\mu\to i}\right)_{t}}{\left(\bar{\Sigma}_{\mu\to i}\right)_{tt}}\right)\right)\label{eq:B3}
\end{equation}
Equations~\ref{eq:m1},\ref{eq:n2},\ref{eq:a1},\ref{eq:Sigma1}
and~\ref{eq:B3} form a closed system which allows computations to
be performed effectively, and which can conveniently be modified to derive
a heuristic solver algorithm (see section~\ref{sub:reinBP}). However, as stated at the
beginning of this section, these equations fail to exactly reproduce the replica
theory results.

\paragraph{Corrected Belief Propagation algorithm\label{par:Corrected-BP}}

The reason for the failure of the standard BP equations to provide correct
results (e.g.~when computing the entropy) is that when the inputs are unbalanced,
i.e.~they don't average to $0$ (as is necessarily the case when the values are
in $\left\{ 0,1\right\} $), the clustering property, i.e.~the assumption that that
the messages incoming into variable nodes from different factor nodes are uncorrelated,
is violated. This can be seen by considering (see~\cite{mezard_space_1989}):
\begin{eqnarray*}
c_{\mu\nu\to i} & = & \frac{1}{N}\left(\left\langle \left(\bar{a}_{\mu\to i}\right)_{t}\left(\bar{a}_{\nu\to i}\right)_{t}\right\rangle -\left\langle \left(\bar{a}_{\mu\to i}\right)_{t}\right\rangle \left\langle \left(\bar{a}_{\nu\to i}\right)_{t}\right\rangle \right)\\
 & = & \frac{1}{N}\left(\left\langle \left(\sum_{j\ne i}\xi_{jt}^{\mu}J_{j}\right)\left(\sum_{j\ne i}\xi_{jt}^{\nu}J_{j}\right)\right\rangle -\left\langle \sum_{j\ne i}\xi_{jt}^{\mu}J_{j}\right\rangle \left\langle \sum_{j\ne i}\xi_{jt}^{\nu}J_{j}\right\rangle \right)\\
 & = & \frac{1}{N}\sum_{j\ne i}\xi_{jt}^{\mu}\xi_{jt}^{\nu}\left(1-m_{j\to\mu}m_{j\to\nu}\right)
\end{eqnarray*}
which is $\mathcal{O}\left(1\right)$ unless the average input $\bar{\xi}$
is zero, in which case it is $\mathcal{O}\left(N^{-\frac{1}{2}}\right)$
and becomes negligible. Only in that case, therefore, standard BP equations
become asymptotically correct for large $N$; in all other circumstances, 
they only provide an approximation (numerical experiments show that for our
model they sistematically predict a slightly lower entropy than the correct one).

We also note that, if we define $\xi_{it}^{\mu}=\bar{\xi}+\rho_{it}^{\mu}$,
where $\bar{\xi}=f$ and $\rho_{it}^{\mu}\in\left\{ -f,1-f\right\} $
with average $\bar{\rho}=0$, we can split the depolarization as such:
\begin{equation}
\sum_{i}J_{i}\xi_{it}^{\mu} = f\sum_{i}J_{i}+\sum_{i}J_{i}\rho_{it}^{\mu}\equiv f\sqrt{N}T+\sum_{i}J_{i}\rho_{it}^{\mu}
\end{equation}
where we defined the overall magnetization $T=\frac{1}{\sqrt{N}}\sum_{i}J_{i}$.
It becomes apparent that the depolarization distributions induced
by the different patterns are all correlated via $T$, which is a
global quantity. We can however amend the BP algorithm, and derive
correct marginals and therefore correct global thermodynamic quantities,
by studying a related problem, in which this contribution is removed from the
factor nodes and induced by an external field instead; this suggests
the following modification to the cavity equations: we start by choosing
a value for the magnetization, call it $T^{\prime}$, and we consider the
problem with patterns $\rho$ instead of $\xi$, thereby ensuring
that the clustering property holds, and with an additional external field $F$
applied to each variable node, thus modifying eqs.~\ref{eq:m1}
and~\ref{eq:n1} as such:
\begin{eqnarray}
m_{i\to\mu} & = & \tanh\left(\sum_{\nu\ne\mu}\tanh^{-1}\left(n_{\nu\to i}\right) + F\right)\label{eq:m_corr}\\
n_{\mu\to i} & \propto & P\left(\sigma_{\textrm{exp}}^{\mu}=1-\prod_{t=1}^{K}\Theta\left(\theta-\sum_{j\ne i}J_{j}\rho_{jt}^{\mu}-\rho_{it}^{\mu}\right)\right)+\label{eq:n_corr}\\
 &  & -P\left(\sigma_{\textrm{exp}}^{\mu}=1-\prod_{t=1}^{K}\Theta\left(\theta-\sum_{j\ne i}J_{j}\rho_{jt}^{\mu}+\rho_{it}^{\mu}\right)\right)\nonumber 
\end{eqnarray}
The total magnetization $T$ can be obtained from the cavity marginals as:
\begin{equation}
T = \sum_i\tanh\left(\sum_{\mu}\tanh^{-1}\left(n_{\mu\to i}\right) + F\right)\label{eq:T}
\end{equation}
Therefore, we can ensure that, at the fixed point, $T=T^{\prime}$ by just
adding an extra step to the BP iterative process in which $F$ is modified
at each iteration according to the difference $T-T^\prime$.

\begin{figure}
\includegraphics{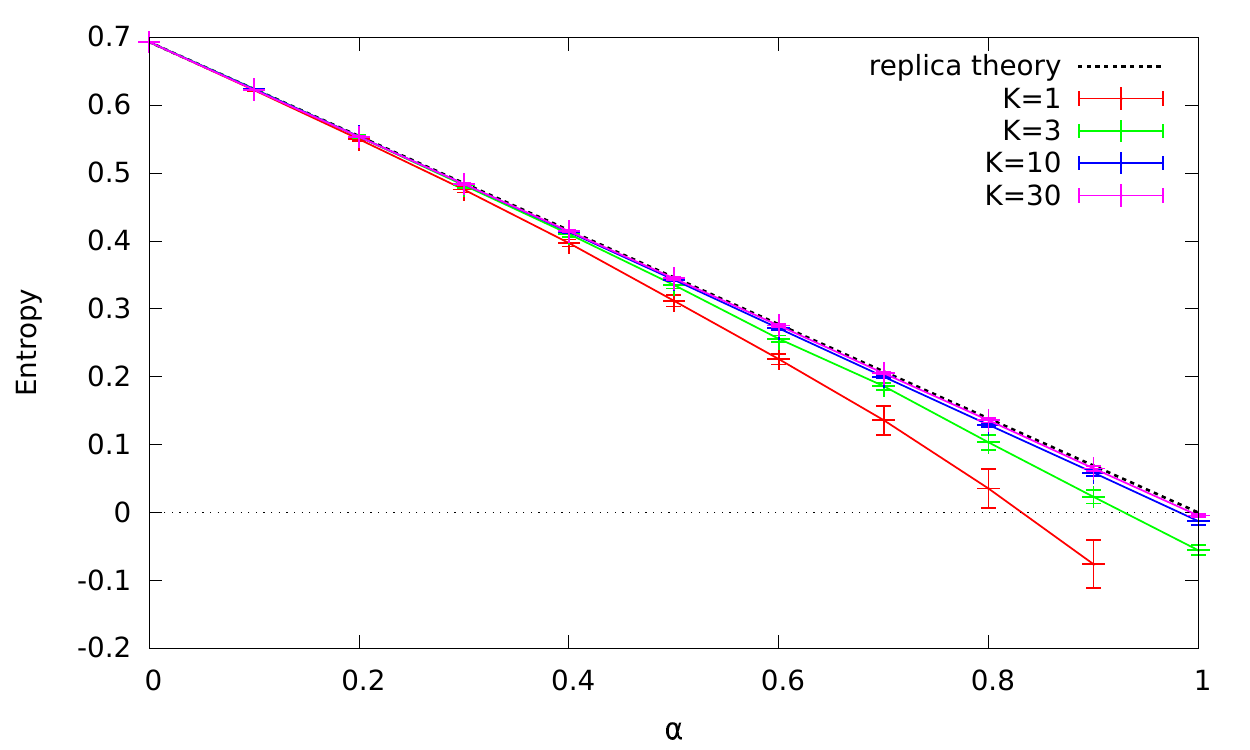}\caption{\label{fig:entropy}Entropy vs. $\alpha$ as computed by the cavity
method at different values of $K$, compared to the one predicted
by the replica theory for $K\gg1$ and $N\gg K$. Each point shows
the average and standard deviation over $10$ random samples with
$N=1000$.}
\end{figure}

After convergence, we compute the entropy $S\left(T^{\prime}\right)$,
and via this define $T^{\star}=\textrm{argmax}_{T^{\prime}}S\left(T^{\prime}\right)$.
Then, the marginals computed for the problem defined by $T^{\star}$
are the same as those to the original problem, and are asymptotically
correct (within the RS assumption), allowing us to compute all the
desired properties on a given instance of the original problem via
this modified cavity method. By averaging over many different samples,
we can recover the results of the replica method, as shown for the
entropy curves in fig.~\ref{fig:entropy}.

\section{Solving single instances: learning algorithms}

\subsection{Reinforced Belief Propagation\label{sub:reinBP}}

Belief propagation equations, in their message passing form over single
problem instances, have repeatedly proven to provide very effective
heuristics when modified
in order to produce optimal configurations \cite{braunstein_encoding_2007,baldassi_efficient_2007,bailly-bechet_finding_2010}.
Two main ways in which this can be achieved are decimation \cite{mezard_analytic_2002,braunstein_survey_2005}
and reinforcement \cite{braunstein_learning_2006}, which can be seen
as a ``soft decimation'' process. In decimation, cycles are performed
which alternate message passing and fixing (or ``freezing'', or
``decimating'') the most polarized free variables, until all variables
are fixed. In reinforcement, the iterative equations have an additional
term which has the role of a time-dependent external field, and which
is computed from the magnetizations obtained at the preceding time
step, so that the system is driven towards a completely polarized
state. Following \cite{braunstein_learning_2006} the reinforced BP
equations are the same as the normal BP equations with one single
difference for eq.~\ref{eq:m1}, which becomes:
\begin{equation}
m_{i\to\mu}^{\tau+1}=\tanh\left(\gamma\left(\tau\right)\tanh^{-1}\left(m_{i}^{\tau}\right)+\sum_{\nu\ne\mu}\tanh^{-1}\left(n_{\nu\to i}^{\tau+1}\right)\right)\label{eq:m2_reinf}
\end{equation}
where $\tau$ is the iteration step, $m_{i}^{\tau}=\tanh\left(\sum_{\nu}\tanh^{-1}\left(n_{\nu\to i}^{\tau}\right)\right)$
is the total magnetization of variable $i$ at iteration $\tau$,
and $\gamma\left(\tau\right)$ is an iteration-dependent reinforcement
parameter, which we set as $\gamma\left(\tau\right)=\gamma_{0}\tau$.
Note that the additional term is proportional to the iteration step,
and therefore dominates for large $\tau$, ensuring that polarization
towards one single configuration is eventually achieved (although
in practice computational problems will arise in difficult or unsatisfiable
situations, due to the limited precision of the floating point representation).

We implemented both the decimation and the reinforcement schemes,
for both the standard version of BP (for which the marginals are
approximate due to correlations between different messages) and the
corrected version (for which marginals are exact, at the cost of increased
computational complexity and running time). Since we did not find
the corrected version to provide any significant advantage over the
na\"ive version (which is not particularly surprising, considering that
the approximation provided by standard BP is rather good, and that the introduction
of the reinforcement term introduces spurious correlations by itself),
here we will only present results for the latter case.

The value of $\gamma_{0}$ is a parameter of the solver algorithm;
higher values of $\gamma_{0}$ make the algorithm greedier, in that
the messages are polarized more quickly but can get trapped into a
non-zero-energy state, while reducing $\gamma_{0}$ improves the accuracy
of the algorithm at the cost of requiring more iterations. In practical
tests, we found that by using eq.~\ref{eq:n2} we were able to reach
values of $\alpha$ as high as $0.7$, but only at the cost of using
extremely small values of $\gamma_{0}$ (of the order of $10^{-7}$
for $N=1000$ and $K=10$), and therefore of an impractically high
computational time.

However, we found heuristically that, by detecting when an excessively
polarized state was reached, and introducing a ``depolarization event''
triggered by such condition, we could achieve the same results with
much higher values of $\gamma_{0}$, and therefore in a much shorter
computational time. More in detail, we introduced, at each iteration
step, a check to detect cases in which any of the terms in the denominator
if eq.~\ref{eq:n2} goes to $0$, indicating a numerical problem
due to excessively polarized magnetizations in a state of non-zero
energy. Whenever this condition is found, we divide all messages $m_{i\to\mu}^{\tau}$
and total magnetizations $m_{i}^{\tau}$ by a positive factor $b$
(thus depolarizing all the messages), and reset $\gamma\left(\tau\right)$
to $0$. In subsequent iterations, we keep increasing $\gamma\left(\tau\right)$
linearly in steps of $\gamma_{0}$ (until another event is detected).
We obtain good results by setting the factor $b$ to $2$ initially,
and increasing it by one at every invocation of this additional depolarization
rule. Indeed, since $\gamma\left(\tau\right)$ does not increase monotonically
any more in this scheme, this modified algorithm will not be guaranteed
to polarize towards a single state, unless the state itself has zero
energy and therefore represents a solution to the problem.

\begin{figure}
\includegraphics[scale=0.6]{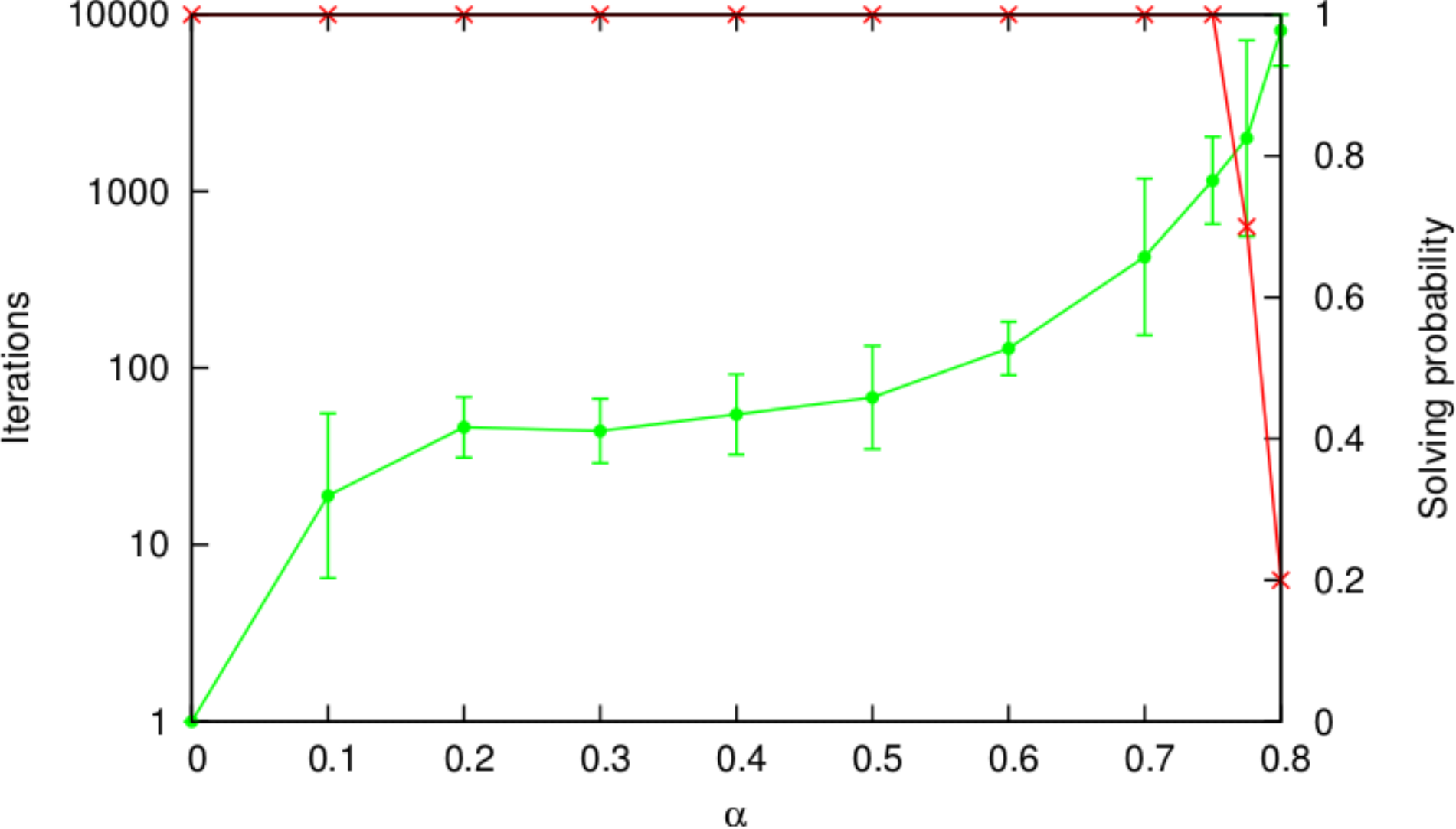}

\caption{\label{fig:bprein}Number of iterations until solving (green curve)
and solving probability (red curve) for different values of $\alpha$,
for a tempotron device with $N=1000$ and $K=10$, with the reinforced
BP algorithm. The parameter $\gamma$ was set to $0.01$. For each
point, $10$ samples were used. The number of iterations was capped
at $10000$.}
\end{figure}

Fig.~\ref{fig:bprein} shows the performance of this algorithm for
$N=1000$ and $K=10$. Setting $10000$ as the maximum number of iterations,
a critical capacity of almost $0.8$ is achieved.

\subsection{Simplified BP-inspired scheme\label{sub:BPI}}

As for the case of the simple perceptron \cite{braunstein_learning_2006,baldassi_efficient_2007,baldassi_generalization_2009},
it is possible to drastically simplify the reinforced BP equations
(in a purely heuristic way), and obtain an online algorithm which,
when parameters are set to their optimal values, proves to be almost
as effective at learning as reinforced BP itself, while dramatically
reducing computational requirements.

This algorithm requires a hidden state $h_{i}$ to be endowed with
each synapse. This hidden state can only assume odd integer values,
and is capped by a maximum absolute value $h^{\max}$, so that each
synapse has a total of $h^{\max}+1$ hidden states. The hidden state
and the synaptic weight $J_{i}$ are related by the simple expression
$J_{i}=\textrm{sign}\left(h_{i}\right)$. In all simulations, we set
the initial state of the $h_{i}$ states by randomly drawing values from
$\left\{ -1,1\right\} $.

The learning protocol turns out to be as follows: patterns $\xi^{\mu}$
are presented in random order to the device, computing the depolarization
$\Delta_{t}^{\mu}=\left(\sum_{i=1}^{N}J_{i}\xi_{it}^{\mu}-\theta\right)$;
from this, we determine $t^{\star}=\textrm{argmax}_{t}\Delta_{t}^{\mu}$
and compute $\Phi^{\mu}=\sigma_{\exp}^{\mu}\Delta_{t^{\star}}^{\mu}$;
depending on the value of $\Phi^{\mu}$, we choose one of three actions:
\begin{description}
\item [{$\Phi^{\mu}>1$}]: do nothing
\item [{$0<\Phi^{\mu}\le1$}]: with probability $r$, update synapses for
which $\xi_{it^{\star}}^{\mu}=1$ and $J_{i}=\sigma_{\exp}^{\mu}$;
with probability $\left(1-r\right)$ do nothing
\item [{$\Phi^{\mu}\le0$}]: update all synapses for which $\xi_{it^{\star}}^{\mu}=1$
\end{description}
The synaptic update rule is always of this form:
\[
h_{i}\to h_{i}+2\sigma_{\exp}^{\mu}
\]
which implies that synaptic values $J_{i}$ are only updated in the
$\Phi^{\mu}<0$ case, and only if $\xi_{it^{\star}}^{\mu}=1$ and
$h_{i}=-\sigma_{\exp}^{\mu}$. As stated above, we impose $-h^{\max}\le h_{i}\le h^{\max}$,
so that the update rule is not applied when $h_{i}=h^{\max}\sigma_{\exp}^{\mu}$.

\begin{figure}
\includegraphics[scale=0.6]{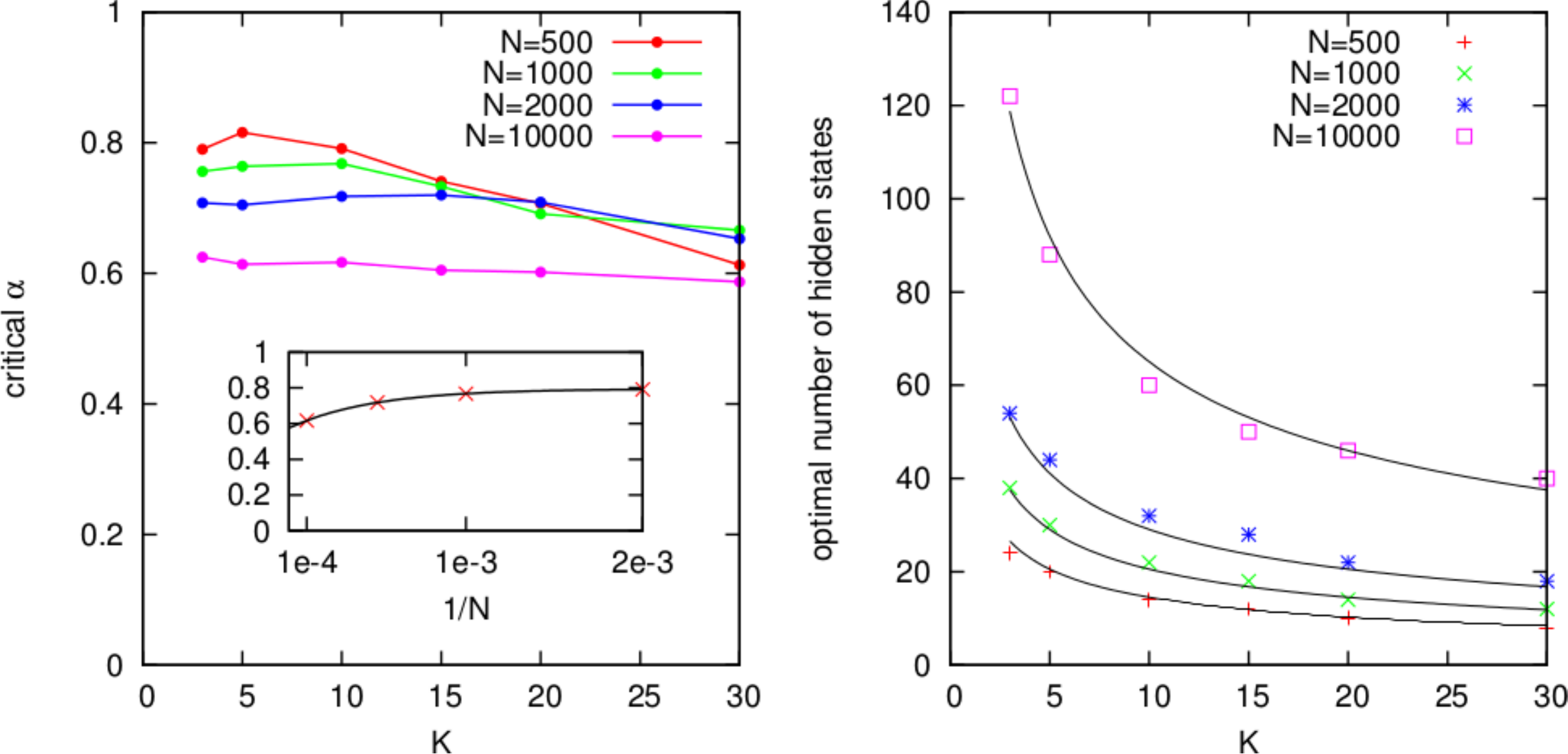}

\caption{\label{fig:simplestats}Optimal critical capacity (left) and optimal
value of the number of hidden states $h^{\max}+1$ (right) for various
values of $N$ and $K$. Critical capacity is defined as the value
of $\alpha$ for which the probability of successfully solving the
problem in $10000$ iterations or less is $0.5$. Optimal $h^{\max}$
is defined as the value of $h^{\max}$ which yields the highest critical
capacity. The parameter $r$ is set to $0.4$ in all plots shown here,
since that was found to be the optimal value independently of other
parameters. At least $40$ random samples were generated for each
combination of $\left(N,\alpha,K,h^{\max},r\right)$ in order to determine
the success probability and therefore the critical capacity. $\alpha$
was explored in steps of $0.05$. The inset in the left panel shows
the critical capacity as a function of $\nicefrac{1}{N}$ for $K=10$;
the solid black line is a fit by an exponential function. The solid
black lines in the right panel show the fit of $h^{\max}+1$ as a
function of $N$ and $K$ via $\lambda\sqrt{\nicefrac{N}{K}}$.}
\end{figure}

The probability $r$ of taking an action in the ``barely correct''
case $0\le\Phi^{\mu}\le1$ is a parameter of the algorithm, just as
$h^{\max}$. We determined empirically the optimal values of $r$
and $h^{\max}$ for different values of $N$ and $K$ by extensively
testing the space of the parameters. Our results, shown in fig.~\ref{fig:simplestats},
indicate that $r=0.4$ works best for all values (we explored the
values of $r$ in steps of $0.05)$, and that the optimal value of
$h^{\max}$ is reasonably well fitted by a function $\lambda\sqrt{\nicefrac{N}{K}}$,
where $\lambda=2.06\pm0.02$. The capacity decreases with $N$, for
fixed $K$, but does not seem to tend to $0$ asymptotically (see
inset in fig.~\ref{fig:simplestats}).

\subsection{Generality of the discrete-time model\label{sub:Continuous}}

As a way to verify that the model and learning protocols which we
studied are relevant in a more biologically realistic setting, we
adapted the simplified BP-inspired scheme described in the previous
section to address the continuous-time classification problem (see
Introduction): for a given an instance of the problem, we discretize
the time in $K$ bins, apply the BP-inspired learning protocol
(slightly modified to use continuous inputs), and test the resulting
synaptic weights assignments on the continuous device (see the
Appendix for details,
section~\ref{sec:Appendix:-time-discretization}). We found that in a
device with $N=1000$ synapses, with time constants $t_m=10ms$ and
$t_s=2.5ms$, tested on $T=500ms$ long input patterns discretized in
$50$ bins, this scheme can achieve a classification error lower than
1\% up to $\alpha=0.4$, demonstatrating that indeed under these
conditions not much relevant information is typically lost in the time
discretization process, and that the proposed time-discretized
learning protocol can be effective even in a continuous-time setting.

\section{Conclusions}

We have presented a theoretical analysis of the computational
performance of the tempotron model with discretized time and discrete
synaptic weights.  The results show that the device is able to learn
random spatio-temporal patterns at a learning rate which saturates the
information theoretic bounds.

In addition to this, and possibly of more practical interest, we have
been able to derive some novel learning protocols which are local
and distributed and do not rely on a gradient descent process on the synaptic
weights. These algorithms are based on the message-passing method
and extend previous works on rate-coding networks. Specifically, we
have shown that the message-passing algorithms can store spatio-temporal
patterns at very high loads and that even some extremely simplified
versions are still able to store an extensive number of patterns
efficiently. Furthermore, we showed that the discretized-time
algorithm can even be adapted to effectively address the original,
continuous-time version of the problem. Our approach can be applied to
both discrete and continuous synapses.

Many open problems remain to be studied, starting from how these protocols
can be made even simpler in a biologically plausible modeling context.
Still we believe that at least as far as artificial neural systems
is concerned these results could find direct application in neuromorphic
devices.

\section{Appendix: statistical physics analysis and replica calculations\label{sec:Appendix:-replica-calculations}}

\subsection{Entropy\label{sub:Entropy}}

We will consider the case in which synaptic weights take values in
$\left\{ -1,1\right\} $ first.

The volume of the space of the solutions for a given instantiation
of the patterns can be written as:
\begin{equation}
\mathcal{V}=\sum_{\left\{ J_{i}\right\} _{i}}\sum_{\left\{ \tau_{t}^{\mu}\right\} _{\mu t}}\prod_{\mu}\chi\left(s_{\textrm{exp}}^{\mu},\left\{ \tau_{t}^{\mu}\right\} _{t}\right)\prod_{\mu t}\Theta\left(\tau_{t}^{\mu}\left(\frac{1}{\sqrt{N}}\sum_{i}J_{i}\xi_{it}^{\mu}-\frac{\theta}{\sqrt{N}}\right)\right)\label{eq:V}
\end{equation}
where index $i\in\left\{ 1,\dots,N\right\} $ is used for synapses,
index $t\in\left\{ 1,\dots,K\right\} $ is used for time bins, index
$\mu\in\left\{ 1,\dots,\alpha N\right\} $ is used for patterns, the
auxiliary variables $\tau_{t}^{\mu}\in\left\{ -1,1\right\} $ are
the internal representations (they are equal to $2V_{t}^{\mu}-1$,
see section~\ref{sec:Introduction}), and$\chi\left(s,\left\{ \tau_{t}\right\} _{t}\right)=\Theta\left(s-\left(1-\prod_{t=1}^{K}\left(1-\frac{1}{2}\left(1+\tau_{t}^{\mu}\right)\right)\right)\right)$
is a characteristic function ensuring that the internal representation
$\tau_{t}$ is compatible with the output $s$.

From here on, for simplicity, we will omit the subscript $\textrm{exp}$
from the outputs $s^{\mu}$.

In order to compute the entropy, we need to compute the quenched average
$\left\langle \log\mathcal{V}\right\rangle _{\xi,s}$; we do this
by using the replica trick:\cite{braunstein_learning_2006,baldassi_efficient_2007}
\begin{equation}
\left\langle \log\mathcal{V}\right\rangle _{\xi,s}=\lim_{n\to0}\frac{\left\langle \mathcal{V}^{n}\right\rangle _{\xi,s}-1}{n}
\end{equation}

where we compute $\left\langle \mathcal{V}^{n}\right\rangle _{\xi,s}$
for integer values of $n$, and use the analytic continuation to compute
the limit $n\to0$. The average over the replicated volume is:
\begin{equation}
\left\langle \mathcal{V}^{n}\right\rangle _{\xi,s}=\left\langle \sum_{\left\{ J_{i}^{a}\right\} _{ia}}\sum_{\left\{ \tau_{t}^{\mu a}\right\} _{\mu ta}}\prod_{\mu a}\chi\left(s^{\mu},\left\{ \tau_{t}^{\mu a}\right\} _{t}\right)\prod_{\mu ta}\Theta\left(\tau_{t}^{\mu a}\left(\frac{1}{\sqrt{N}}\sum_{i}J_{i}^{a}\xi_{it}^{\mu}-\frac{\theta}{\sqrt{N}}\right)\right)\right\rangle _{\xi,s}\label{eq:Vol^n}
\end{equation}
We used the index $a\in\left\{ 1,\dots,n\right\} $ to denote the
replica. We can now use the integral representation of the $\Theta$
function, $\Theta\left(y\right)=\int_{-\infty}^{\infty}\frac{dx}{2\pi}\int_{0}^{\infty}d\lambda\, e^{ix\left(\lambda-y\right)}$,
and compute the average over the input patterns, using their independence
and the $N\gg1$ limit (in the following, all integrals are assumed
to be on $\left[-\infty,\infty\right]$ unless otherwise specified)
:
\begin{eqnarray}
 &  & \left\langle \prod_{a}\Theta\left(\tau_{t}^{\mu a}\left(\frac{1}{\sqrt{N}}\sum_{i}J_{i}^{a}\xi_{it}^{\mu}-\frac{\theta}{\sqrt{N}}\right)\right)\right\rangle _{\xi}=\\
 &  & \quad=\int\prod_{a}\frac{dx_{t}^{\mu a}}{2\pi}\int_{0}^{\infty}\prod_{a}d\lambda_{t}^{\mu a}\,\prod_{a}\exp\left(ix_{t}^{\mu a}\left(\lambda_{t}^{\mu a}-\tau_{t}^{\mu a}\left(\bar{\xi}\frac{1}{\sqrt{N}}\sum_{i}J_{i}^{a}-\frac{\theta}{\sqrt{N}}\right)\right)\right)\cdot\nonumber \\
 &  & \quad\qquad\cdot\exp\left(-\frac{v_{\xi}}{2N}\sum_{a,b}\tau_{t}^{\mu a}\tau_{t}^{\mu b}x_{t}^{\mu a}x_{t}^{\mu b}\sum_{i}J_{i}^{a}J_{i}^{b}\right)\nonumber 
\end{eqnarray}
where $\bar{\xi}=f$ and $v_{\xi}=f\left(1-f\right)$ are the average
value and the variance of the inputs $\xi_{it}^{\mu}$, respectively.
We then introduce order parameters $q^{ab}=\frac{1}{N}\sum_{i}J_{i}^{a}J_{i}^{b}$
and $T^{a}=-\sqrt{N}\bar{J}+\frac{1}{\sqrt{N}}\sum_{i}J_{i}^{a}$
via Dirac-delta functions, and their conjugates $\hat{q}^{ab}$,$\hat{T}^{a}$
via integral expansion of the deltas, and get:
\begin{eqnarray}
\left\langle \mathcal{V}^{n}\right\rangle _{\xi,s} & = & \int\prod_{a}\frac{dT^{a}d\hat{T^{a}}\sqrt{N}}{2\pi}\int\prod_{a\ge b}\frac{dq^{ab}d\hat{q}^{ab}N}{2\pi}\exp\left(\sqrt{N}\sum_{a}\left(T^{a}+\sqrt{N}\bar{J}\right)\hat{T}^{a}-N\sum_{a\ge b}q^{ab}\hat{q}^{ab}\right)\cdot\\
 &  & \cdot\left(\sum_{\left\{ J_{i}^{a}\right\} _{ia}}\prod_{i}\exp\left(\sum_{a\ge b}\hat{q}^{ab}J_{i}^{a}J_{i}^{b}-\sum_{a}\hat{T}^{a}J_{i}^{a}\right)\right)\cdot\nonumber \\
 &  & \cdot\left\langle \sum_{\left\{ \tau_{t}^{\mu a}\right\} _{\mu ta}}\prod_{\mu a}\chi\left(s^{\mu},\left\{ \tau_{t}^{\mu a}\right\} _{t}\right)\right.\cdot\nonumber \\
 &  & \quad\cdot\prod_{\mu t}\left(\int\prod_{a}\frac{dx_{t}^{\mu a}}{2\pi}\int_{0}^{\infty}\prod_{a}d\lambda_{t}^{\mu a}\,\prod_{a}\exp\left(ix_{t}^{\mu a}\left(\lambda_{t}^{\mu a}-\tau_{t}^{\mu a}\left(\bar{\xi}\left(T^{a}+\sqrt{N}\bar{J}\right)-\frac{\theta}{\sqrt{N}}\right)\right)\right)\right.\cdot\nonumber \\
 &  & \quad\left.\left.\cdot\exp\left(-\frac{v_{\xi}}{2}\sum_{a,b}\tau_{t}^{\mu a}\tau_{t}^{\mu b}x_{t}^{\mu a}x_{t}^{\mu b}q^{ab}\right)\right)\right\rangle _{s}\nonumber \\
 & = & \int\prod_{a}\frac{dT^{a}d\hat{T^{a}}\sqrt{N}}{2\pi}\int\prod_{a\ge b}\frac{dq^{ab}d\hat{q}^{ab}N}{2\pi}\exp\left(\sqrt{N}\sum_{a}\left(T^{a}+\sqrt{N}\bar{J}\right)\hat{T}^{a}-N\sum_{a\ge b}q^{ab}\hat{q}^{ab}\right)\cdot\nonumber \\
 &  & \cdot\left(\sum_{\left\{ J^{a}\right\} _{a}}\exp\left(\sum_{a\ge b}\hat{q}^{ab}J^{a}J^{b}-\sum_{a}\hat{T}^{a}J^{a}\right)\right)^{N}\cdot\nonumber \\
 &  & \cdot\left\langle \sum_{\left\{ \tau_{t}^{a}\right\} _{ta}}\prod_{a}\chi\left(s,\left\{ \tau_{t}^{a}\right\} _{t}\right)\right.\cdot\nonumber \\
 &  & \quad\cdot\prod_{t}\left(\int\prod_{a}\frac{dx_{t}^{a}}{2\pi}\int_{0}^{\infty}\prod_{a}d\lambda_{t}^{a}\,\prod_{a}\exp\left(ix_{t}^{a}\left(\lambda_{t}^{a}-\tau_{t}^{a}\left(\bar{\xi}\left(T^{a}+\sqrt{N}\bar{J}\right)-\frac{\theta}{\sqrt{N}}\right)\right)\right)\right.\cdot\nonumber \\
 &  & \quad\left.\left.\cdot\exp\left(-\frac{v_{\xi}}{2}\sum_{a,b}\tau_{t}^{a}\tau_{t}^{b}x_{t}^{a}x_{t}^{b}q^{ab}\right)\right)\right\rangle _{s}^{\alpha N}\nonumber 
\end{eqnarray}
where in the second step we dropped indices $i$ and $\mu$. We expand
the threshold $\theta$ in series of $\sqrt{N}$:
\begin{equation}
\theta=N\theta_{0}+\sqrt{N}\theta_{1}
\end{equation}
from which we immediately get the relation:
\begin{equation}
\bar{J}=\frac{\theta_{0}}{\bar{\xi}}
\end{equation}

This leaves us with:
\begin{eqnarray}
\left\langle \mathcal{V}^{n}\right\rangle _{\xi,s} & = & \int\prod_{a}\frac{dT^{a}d\hat{T^{a}}\sqrt{N}}{2\pi}\int\prod_{a\ge b}\frac{dq^{ab}d\hat{q}^{ab}N}{2\pi}\cdot\\
 &  & \cdot\exp\left(\sqrt{N}\sum_{a}T^{a}\hat{T}^{a}+N\bar{J}\sum_{a}\hat{T}^{a}-N\sum_{a\ge b}q^{ab}\hat{q}^{ab}\right)\cdot\nonumber \\
 &  & \cdot\left(\sum_{\left\{ J^{a}\right\} _{a}}\exp\left(\sum_{a\ge b}\hat{q}^{ab}J^{a}J^{b}-\sum_{a}\hat{T}^{a}J^{a}\right)\right)^{N}\cdot\nonumber \\
 &  & \cdot\left\langle \sum_{\left\{ \tau_{t}^{a}\right\} _{ta}}\prod_{a}\chi\left(s,\left\{ \tau_{t}^{a}\right\} _{t}\right)\prod_{t}\left(\int\prod_{a}\frac{dx_{t}^{a}}{2\pi}\int_{0}^{\infty}\prod_{a}d\lambda_{t}^{a}\,\prod_{a}\exp\left(ix_{t}^{a}\left(\lambda_{t}^{a}-\tau_{t}^{a}\left(\bar{\xi}T^{a}-\theta_{1}\right)\right)\right)\right.\right.\cdot\nonumber \\
 &  & \quad\left.\left.\cdot\exp\left(-\frac{v_{\xi}}{2}\sum_{a,b}\tau_{t}^{a}\tau_{t}^{b}x_{t}^{a}x_{t}^{b}q^{ab}\right)\right)\right\rangle _{s}^{\alpha N}\nonumber 
\end{eqnarray}

In the $N\gg1$ limit, this integral can be computed by the saddle
point method: we introduce the RS Ansatz for the solution: $T^{a}=T\,\forall a$,
$q^{ab}=q\,\forall a,b:a\ne b$, $q^{aa}=Q\,\forall a$, and analogous
expressions for the conjugate parameters. Therefore:
\begin{eqnarray}
\left\langle \mathcal{V}^{n}\right\rangle _{\xi,s} & = & \exp\left(N\bar{J}\hat{T}+N\frac{n}{2}\hat{q}q-Nn\hat{Q}Q\right)\cdot\\
 &  & \cdot\left(\sum_{\left\{ J^{a}\right\} _{a}}\exp\left(\frac{\hat{q}}{2}\left(\sum_{a}J^{a}\right)^{2}-\frac{1}{2}\left(\hat{q}-2\hat{Q}\right)\sum_{a}\left(J^{a}\right)^{2}-\hat{T}\sum_{a}J^{a}\right)\right)^{N}\cdot\nonumber \\
 &  & \cdot\left\langle \sum_{\left\{ \tau_{t}^{a}\right\} _{ta}}\prod_{a}\chi\left(s,\left\{ \tau_{t}^{a}\right\} _{t}\right)\prod_{t}\left(\int\prod_{a}\frac{dx_{t}^{a}}{2\pi}\int_{0}^{\infty}\prod_{a}d\lambda_{t}^{a}\,\prod_{a}\exp\left(ix_{t}^{a}\left(\lambda_{t}^{a}-\tau_{t}^{a}\left(\bar{\xi}T-\theta_{1}\right)\right)\right)\right.\right.\cdot\nonumber \\
 &  & \quad\left.\left.\cdot\exp\left(-\frac{v_{\xi}}{2}\left(q\left(\sum_{a}\tau_{t}^{a}x_{t}^{a}\right)^{2}+\left(Q-q\right)\sum_{a}\left(x_{t}^{a}\right)^{2}\right)\right)\right)\right\rangle _{s}^{\alpha N}\nonumber \\
 & = & \exp\left(N\bar{J}\hat{T}+N\frac{n}{2}\hat{q}q-Nn\hat{Q}Q\right)\cdot\nonumber \\
 &  & \cdot\left(\int Du\left(\sum_{\left\{ J\right\} }\exp\left(-\frac{1}{2}\left(\hat{q}-2\hat{Q}\right)J^{2}+\left(\sqrt{\hat{q}}u-\hat{T}\right)J\right)\right)^{n}\right)^{N}\cdot\nonumber \\
 &  & \cdot\left(\int\prod_{t}Du_{t}\left\langle \left(\sum_{\left\{ \tau_{t}\right\} _{t}}\chi\left(s,\left\{ \tau_{t}\right\} _{t}\right)\prod_{t}\left(\int\frac{dx_{t}}{2\pi}\int_{0}^{\infty}d\lambda_{t}\,\exp\left(-\frac{v_{\xi}}{2}\left(Q-q\right)\left(x_{t}\right)^{2}\right)\right.\right.\right.\right.\cdot\nonumber \\
 &  & \quad\left.\left.\left.\left.\cdot\exp\left(ix_{t}\left(\lambda_{t}-\tau_{t}\left(\bar{\xi}T-\theta_{1}-\sqrt{v_{\xi}q}u_{t}\right)\right)\right)\right)\right)^{n}\right\rangle _{s}\right)^{\alpha N}\nonumber \\
 & = & \exp\, Nn\left(\bar{J}\hat{T}+\frac{1}{2}\hat{q}q-\hat{Q}Q\right.+\nonumber \\
 &  & \qquad+\int Du\,\log\left(\sum_{\left\{ J\right\} }\exp\left(-\frac{1}{2}\left(\hat{q}-2\hat{Q}\right)J^{2}+\left(\sqrt{\hat{q}}u-\hat{T}\right)J\right)\right)\nonumber \\
 &  & \qquad+\left.\alpha\int\prod_{t}Du_{t}\left\langle \log\left(\sum_{\left\{ \tau_{t}\right\} _{t}}\chi\left(s,\left\{ \tau_{t}\right\} _{t}\right)\prod_{t}H\left(-\tau_{t}\frac{\bar{\xi}T-\theta_{1}-\sqrt{v_{\xi}q}u_{t}}{\sqrt{v_{\xi}\left(Q-q\right)}}\right)\right)\right\rangle _{s}\right)\nonumber 
\end{eqnarray}
where in the second step we introduced auxiliary Gaussian integrals
(we use the shorthand notation $Du=du\frac{1}{\sqrt{2}\pi}e^{-\frac{u^{2}}{2}}$
and define $H\left(x\right)=\int_x^\infty\ Dy$),
which allows to drop the replica index $a$, and in the last step
we used the $n\to0$ limit. Finally, we obtain the expression for
the entropy:
\begin{eqnarray}
\mathcal{S}=\frac{1}{N}\left\langle \log\mathcal{V}\right\rangle _{\xi,s} & = & \bar{J}\hat{T}+\frac{1}{2}\hat{q}q-\hat{Q}Q+\mathcal{Z}_{J}\left(\hat{Q},\hat{q},\hat{T}\right)+\mathcal{Z}_{S}\left(Q,q,T\right)\\
\mathcal{Z}_{J}\left(\hat{Q},\hat{q},\hat{T}\right) & = & \int Du\,\log\left(\sum_{\left\{ J\right\} }\exp\left(-\frac{1}{2}\left(\hat{q}-2\hat{Q}\right)J^{2}+\left(\sqrt{\hat{q}}u-\hat{T}\right)J\right)\right)\\
\mathcal{Z}_{S}\left(Q,q,T\right) & = & \alpha\int\prod_{t}Du_{t}\left\langle \log\left(\sum_{\left\{ \tau_{t}\right\} _{t}}\chi\left(s,\left\{ \tau_{t}\right\} _{t}\right)\prod_{t}H\left(-\tau_{t}\,\eta\left(u_{t},Q,q,T\right)\right)\right)\right\rangle \\
\eta\left(u,Q,q,T\right) & = & \frac{\bar{\xi}T-\theta_{1}-\sqrt{v_{\xi}q}u}{\sqrt{v_{\xi}\left(Q-q\right)}}\label{eq:eta}
\end{eqnarray}
The expression for $\mathcal{Z}_{J}$ is the familiar expression for
perceptron models, and it can be written more explicitly for the two
cases $J\in\left\{ -1,1\right\} $ and $J\in\left\{ 0,1\right\} $:
\begin{eqnarray}
\mathcal{Z}_{J}^{\pm}\left(\hat{Q},\hat{q},\hat{T}\right) & = & -\frac{1}{2}\left(\hat{q}-2\hat{Q}\right)+\int Du\,\log\left(2\cosh\left(\sqrt{\hat{q}}u-\hat{T}\right)\right)\\
\mathcal{Z}_{J}^{01}\left(\hat{Q},\hat{q},\hat{T}\right) & = & \int Du\,\log\left(1+\exp\left(-\frac{1}{2}\left(\hat{q}-2\hat{Q}\right)+\sqrt{\hat{q}}u-\hat{T}\right)\right)
\end{eqnarray}

The expression for $\mathcal{Z}_{S}$ can be manipulated further:
\begin{eqnarray}
\mathcal{Z}_{S}\left(Q,q,T\right) & = & \alpha\left(1-f^{\prime}\right)K\int Du\log H\left(\eta\left(u,Q,q,T\right)\right)+\\
 &  & +\alpha f^{\prime}\int\prod_{t}Du_{t}\log\left(1-\prod_{t}H\left(\eta\left(u_{t},Q,q,T\right)\right)\right)
\end{eqnarray}

In the limit of $K\gg1$, we can use the central limit theorem and
keep only the higher order terms in $K$, and obtain:
\begin{equation}
\mathcal{Z}_{S}\left(Q,q,T\right)=\alpha\left(\left(1-f^{\prime}\right)K\Lambda\left(Q,q,T\right)+f^{\prime}\log\left(1-\exp\left(K\Lambda\left(Q,q,T\right)\right)\right)\right)
\end{equation}
where we defined:
\begin{equation}
    \Lambda\left(Q,q,T\right)=\int Du\log H\left(\eta\left(u,Q,q,T\right)\right)\label{eq:lambdadef}
\end{equation}

The saddle point equation for $T$ gives:
\[
0=\frac{\partial\mathcal{Z}_{S}}{\partial T}=\alpha K\left(1-f^{\prime}\frac{1}{1-e^{K\Lambda}}\right)\frac{\partial\Lambda\left(Q,q,T\right)}{\partial T}
\]

which implies:
\begin{equation}
\Lambda\left(Q,q,T\right)=\frac{1}{K}\log\left(1-f^{\prime}\right)\label{eq:Lambda}
\end{equation}

This in turn puts to zero $\hat{q}$ and $\hat{Q}$:
\begin{eqnarray*}
\hat{q} & = & -2\frac{\partial\mathcal{Z}_{S}}{\partial q}=-2\alpha K\left(1-f^{\prime}\frac{1}{1-e^{K\Lambda}}\right)\frac{\partial\Lambda\left(Q,q,T\right)}{\partial q}=0\\
\hat{Q} & = & \frac{\partial\mathcal{Z}_{S}}{\partial Q}=\alpha K\left(1-f^{\prime}\frac{1}{1-e^{K\Lambda}}\right)\frac{\partial\Lambda\left(Q,q,T\right)}{\partial Q}=0
\end{eqnarray*}

Optimizing with respect to $\theta_{0}$, i.e.~imposing $\frac{\partial S}{\partial\bar{J}}=0$,
we also get $\hat{T}=0$.

The remaining equations are different for the cases $\pm1$ and $01$.
For the $\pm1$ case:
\begin{eqnarray}
q & = & -2\frac{\partial\mathcal{Z}_{J}}{\partial\hat{q}}=1-\frac{1}{\sqrt{\hat{q}}}\int Du\, u\,\tanh\left(\sqrt{\hat{q}}u-\hat{T}\right)\\
Q & = & \frac{\partial\mathcal{Z}_{J}}{\partial\hat{Q}}=1\\
\bar{J} & = & -\frac{\partial\mathcal{Z}_{J}}{\partial\hat{T}}=\int Du\,\tanh\left(\sqrt{\hat{q}}u-\hat{T}\right)
\end{eqnarray}

The result $Q=1$ is obvious. From $\hat{T}=0$ and $\hat{q}=0$, and
since $\bar{\xi}\neq0$, we get $q=0$ and $\theta_{0}=0$.

For the $01$ case:\cite{braunstein_learning_2006,baldassi_efficient_2007}
\begin{eqnarray}
q & = & -2\frac{\partial\mathcal{Z}_{J}}{\partial\hat{q}}=\int Du\,\frac{1}{1+\exp\left(\frac{1}{2}\left(\hat{q}-2\hat{Q}\right)-\sqrt{\hat{q}}u+\hat{T}\right)}\left(1-\frac{u}{\sqrt{\hat{q}}}\right)\\
Q & = & \frac{\partial\mathcal{Z}_{J}}{\partial\hat{Q}}=\int Du\,\frac{1}{1+\exp\left(\frac{1}{2}\left(\hat{q}-2\hat{Q}\right)-\sqrt{\hat{q}}u+\hat{T}\right)}\\
\bar{J} & = & -\frac{\partial\mathcal{Z}_{J}}{\partial\hat{T}}=Q
\end{eqnarray}

From $\hat{q}=0$ and $\hat{Q}=0$ these simplify to:
\begin{eqnarray*}
Q & = & \bar{J}=\frac{1}{1+e^{\hat{T}}}=\frac{1}{2}\\
q & = & \frac{1}{\left(1+e^{\hat{T}}\right)^{2}}=Q^{2}=\frac{1}{4}\\
\theta_{0} & = & \frac{f}{2}
\end{eqnarray*}

From $q=Q^{2}$ we see that the cross-overlap is as low as possible,
like in the $\pm1$ case: the physical interpretation is that clusters
of solution are isolated, i.e.~point-like.

The only remaining order parameters are $T$ and $\theta_{1}$, which
are related by eq. \ref{eq:Lambda} and give:

\begin{equation}
T=\frac{1}{f}\left(\theta_{1}+\sqrt{2f\left(1-f\right)}\textrm{erfc}^{-1}\left(2\sqrt[K]{1-f^{\prime}}\right)\right)
\end{equation}

Therefore, in order to have unbiased synapses, i.e.~$T=0$, we may
set $\theta_{1}$ to:
\begin{equation}
\theta_{1}=-\sqrt{2f\left(1-f\right)}\textrm{erfc}^{-1}\left(2\sqrt[K]{1-f^{\prime}}\right)\label{eq:theta_1}
\end{equation}
With our choice for the distribution of the inputs, $f=1-\sqrt[K]{1-f}$,
this formula starts from $0$ at $K=1$, has a maximum for $K=6$
and slowly decreases (as $\sqrt{\frac{\log\left(K\right)}{K}}$) to
$0$ as $K$ diverges; the reason for this behaviour is that there
are two competing tendencies at work as $K$ increases: on one hand,
the increase in the length of the internal representation while $f^{\prime}$
is kept constant requires that more and more individual bins fall
below threshold; on the other hand, the sparsification of the inputs
reduces the fluctuations in the depolarization; this second contribution
dominates for large $K$ and so the threshold goes to $0$, but for
practical purposes (i.e.~for biologically relevant values of $K$)
it does not become negligible.

From the above results, we can determine the entropy:
\begin{eqnarray}
\mathcal{S} & = & \log\left(2\right)-\alpha\left(\left(1-f^{\prime}\right)\log\left(1-f^{\prime}\right)+f^{\prime}\log\left(f^{\prime}\right)\right)
\end{eqnarray}
which goes to $0$ at:
\begin{equation}
\alpha_{c}=\left(1-f^{\prime}\right)\log_{2}\left(1-f^{\prime}\right)+f^{\prime}\log_{2}\left(f^{\prime}\right)
\end{equation}
which coincides with the information theoretic upper bound.

\subsection{Distribution of output spikes}

The probability distribution of the number of output spikes (i.e.~$1$'s
in the internal representation) can be obtained by taking the ratio
between the volume of the solution space in which one pattern is restricted
to produce $Y$ spikes and the total volume:
\begin{eqnarray}
P\left(Y\right) & = & \frac{1}{\mathcal{V}}\left\langle \sum_{\left\{ \tau_{t}^{\mu}\right\} _{\mu l}}\delta_{k}\left(\sum_{t}\left(\frac{1+\tau_{t}^{1}}{2}\right),Y\right)\right.\cdot\\
 &  & \quad\cdot\left.\sum_{\left\{ J_{i}\right\} _{i}}\prod_{\mu}\chi\left(s^{\mu},\left\{ \tau_{t}^{\mu}\right\} _{t}\right)\prod_{\mu t}\Theta\left(\tau_{t}^{\mu}\left(\frac{1}{\sqrt{N}}\sum_{i}J_{i}\xi_{it}^{\mu}-\frac{\theta}{\sqrt{N}}\right)\right)\right\rangle _{\xi,s}\nonumber 
\end{eqnarray}
where $\delta_{k}\left(x,y\right)$ is the Kronecker delta function:
\[
\delta_{k}\left(x,y\right)=\begin{cases}
1 & \textrm{if}\, x=y\\
0 & \textrm{otherwise}
\end{cases}
\]

We can write $\mathcal{V}^{-1}=\lim_{n\to0}\mathcal{V}^{n-1}$, restrict
ourselves to integer $n$ and obtain an expression almost identical
to eq.~\ref{eq:Vol^n}, except for the Kronecker delta affecting
pattern $\mu=1$:
\begin{eqnarray}
P\left(Y\right) & = & \left\langle \sum_{\left\{ \tau_{t}^{\mu a}\right\} _{\mu ta}}\delta_{k}\left(\sum_{t}\left(\frac{1+\tau_{t}^{1a}}{2}\right)-Y\right)\right.\cdot\\
 &  & \quad\cdot\left.\sum_{\left\{ J_{i}^{a}\right\} _{ia}}\prod_{\mu a}\chi\left(s^{\mu},\left\{ \tau_{t}^{\mu a}\right\} _{t}\right)\prod_{\mu ta}\Theta\left(\tau_{t}^{\mu a}\left(\frac{1}{\sqrt{N}}\sum_{i}J_{i}^{a}\xi_{it}^{\mu}-\frac{\theta}{\sqrt{N}}\right)\right)\right\rangle _{\xi,s}\nonumber 
\end{eqnarray}

The computation follows the one for the entropy; the only affected
term is $\mathcal{Z}_{S}$, and only one term survives the limit $n\to0$,
giving:
\begin{equation}
P\left(Y\right)=\delta_{k}\left(Y,0\right)\left(1-f^{\prime}\right)+\left(1-\delta_{k}\left(Y,0\right)\right)f^{\prime}\int\prod_{t}Du_{t}\frac{\binom{K}{Y}\prod_{t=1}^{Y}H^{+}\left(u_{t}\right)\prod_{t=Y+1}^{K}H^{-}\left(u_{t}\right)}{1-\prod_{t=1}^{K}H^{-}\left(u_{t}\right)}
\end{equation}
where we wrote $H^{\pm}\left(u\right)=H\left(\mp\eta\left(u,Q,q,t\right)\right)$
(see eq.~\ref{eq:eta}) for short.
For large $K$ and finite $Y$, this approximates to:
\begin{eqnarray}
P\left(Y\right) & = & \delta_{k}\left(Y,0\right)\left(1-f^{\prime}\right)+\left(1-\delta_{k}\left(Y,0\right)\right)f^{\prime}\binom{K}{Y}\frac{e^{\left(K-Y\right)\Lambda}}{1-e^{K\Lambda}}\left(1-e^{\Lambda}\right)^{Y}\\
 & = & \binom{K}{Y}\left(e^{\Lambda}\right)^{K-Y}\left(1-e^{\Lambda}\right)^{Y}\nonumber 
\end{eqnarray}
where $\Lambda=\Lambda\left(Q,q,T\right)$ (see eq.~\ref{eq:lambdadef}), and in the second step
we used eq.~\ref{eq:Lambda} from the saddle point solution. The
result is a binomial distribution, in which the probability of producing
a spike is $1-e^{\Lambda}=1-\sqrt[K]{1-f^{\prime}}$, which is our
choice for the input frequency $f$.

\subsection{Structure of the internal representations}

We can study the structure of the space of the internal representations
by following \cite{monasson_weight_1996,cocco_analytical_1996}: we
consider the volume of each internal representation $\mathcal{V}_{\mathcal{T}}$,
where $\mathcal{T}=\left\{ \tau_{t}^{\mu}\right\} _{\mu t}$ is an
internal representation, such that the overall volume can be written
as $\mathcal{V}=\sum_{\mathcal{T}}\mathcal{V}_{\mathcal{T}}$; then
we define:
\begin{equation}
\mathcal{V}\left(r\right)=\sum_{\mathcal{T}}\left(\mathcal{V_{\mathcal{T}}}\right)^{r}
\end{equation}
and study the free energy defined by:
\begin{equation}
g\left(r\right)=-\frac{\left\langle \log\left(\mathcal{V}\left(r\right)\right)\right\rangle }{Nr}
\end{equation}
which, once known, allows to derive the size of the internal representations
from the quantity 
\begin{equation}
w\left(r\right)=\frac{\partial}{\partial r}\left(-rg\left(r\right)\right)
\end{equation}
 (for $r=1$, $w\left(1\right)=\frac{1}{N}\log\mathcal{V}_{\mathcal{T}}^{\star}$
where $\mathcal{V}_{\mathcal{T}}^{\star}$ is the typical volume of
the dominant internal representations), and their number from the
micro-canonical entropy 
\begin{equation}
\mathcal{N}\left(r\right)=-\frac{\partial}{\partial\nicefrac{1}{r}}g\left(r\right)
\end{equation}
 (for $r=1$, $\mathcal{N}\left(1\right)$ is the logarithm of the
typical number of internal representation of size $\mathcal{V}_{\mathcal{T}}^{\star}$,
divided by $N$).

The computation is performed by using the replica trick for $r$ integer
and then performing an analytic continuation:

\begin{equation}
\left\langle \mathcal{V}\left(r\right)^{n}\right\rangle =\left\langle \sum_{\left\{ \tau_{t}^{\mu a}\right\} _{\mu ta}}\sum_{\left\{ J_{i}^{a\nu}\right\} _{ia\nu}}\prod_{\mu a}\chi\left(s^{\mu},\left\{ \tau_{t}^{\mu a}\right\} _{t}\right)\prod_{\mu ta\nu}\Theta\left(\tau_{t}^{\mu a}\left(\frac{1}{\sqrt{N}}\sum_{i}J_{i}^{a\nu}\xi_{it}^{\mu}-\frac{\theta}{\sqrt{N}}\right)\right)\right\rangle _{\xi,s}
\end{equation}
where we introduced the new internal representation replica index
$\nu\in\left\{ 1,\dots,r\right\} $. The computation follows the steps
of the entropy computation of section~\ref{sub:Entropy}, but requires
the introduction of order parameters with 2 replica indices; in particular
$q^{a\nu,b\phi}=\frac{1}{N}\sum_{i}J_{i}^{a\nu}J_{i}^{b\phi}$, which
in the RS Anzatz can take 3 values:
\[
q^{a\nu,b\phi}=\begin{cases}
Q & \textrm{if}\, a=b,\nu=\phi\\
q_{1} & \textrm{if}\, a=b,\nu\neq\phi\\
q_{0} & \textrm{if}\, a\neq b
\end{cases}
\]
We obtain, for large $K$:
\begin{eqnarray}
g\left(r\right) & = & -\bar{J}\hat{T}-\frac{1}{2}rq_{0}\hat{q}_{0}+\frac{r-1}{2}q_{1}\hat{q}_{1}+\hat{Q}Q-\frac{1}{r}\mathcal{Z}_{J}+\frac{1}{r}\mathcal{Z}_{S}\\
\mathcal{Z}_{J} & = & \int Du\,\log\left(\int Dz\left(\sum_{\left\{ J\right\} }e^{\frac{1}{2}\left(2\hat{Q}-\hat{q}_{1}\right)J^{2}+\left(\sqrt{\hat{q}_{0}}u+\sqrt{\hat{q}_{1}-\hat{q}_{0}}z-\hat{T}\right)J}\right)^{r}\right)\\
\mathcal{Z}_{S} & = & \alpha\left(\left(1-f^{\prime}\right)K\Lambda\left(Q,q_{1},q_{0},T\right)+f^{\prime}\log\left(e^{K\Phi\left(Q,q_{1},q_{0},T\right)}-e^{K\Lambda\left(Q,q_{1},q_{0},T\right)}\right)\right)\\
\Lambda\left(Q,q_{1},q_{0},T\right) & = & \int Du\,\log\left(\int Dz\, H\left(\eta\left(u,z,Q,q_{1},q_{0},T\right)\right)^{r}\right)\\
\Phi\left(Q,q_{1},q_{0},T\right) & = & \int Du\,\log\left(\int Dz\, H\left(-\eta\left(u,z,Q,q_{1},q_{0},T\right)\right)^{r}+H\left(\eta\left(u,z,Q,q_{1},q_{0},T\right)\right)^{r}\right)\hspace{1em}\\
\eta\left(u,z,Q,q_{1},q_{0},T\right) & = & -\frac{u\sqrt{v_{\xi}q_{0}}-z\sqrt{v_{\xi}\left(q_{1}-q_{0}\right)}+\theta_{1}-T\xi}{\sqrt{v_{\xi}\left(Q-q_{1}\right)}}
\end{eqnarray}

The saddle point equations for $r=1$ give the same results as before,
as expected; in particular, we find $q_{0}=q_{1}=0$ in the $\pm1$
case and $q_{0}=q_{1}=\nicefrac{1}{4}$ in the $01$ case, and $g\left(1\right)=-\mathcal{S}$
as expected. Furthermore, we have:
\begin{equation}
\left.\frac{\partial\mathcal{Z_{S}}}{\partial r}\right|_{r=1}=\alpha K\int Du\int Dz\left(H^{+}\left(u,z\right)\log H^{+}\left(u,z\right)+H^{-}\left(u,z\right)\log H^{-}\left(u,z\right)\right)
\end{equation}
where we used the shorthand notation $H^{\pm}\left(u,z\right)=H\left(\mp\eta\left(u,z,Q,q_{1},q_{0},T\right)\right)$.
From this and from the saddle point equations at $r=1$, in particular
from eq.~\ref{eq:Lambda}, we obtain the weight and the entropy of
the dominant internal representations for $f^{\prime}=\nicefrac{1}{2}$:
\begin{eqnarray}
w\left(1\right) & = & \left.\frac{\partial}{\partial r}\left(-rg\left(r\right)\right)\right|_{r=1}=\log2\left(-1+\alpha+\alpha\log K-\alpha\log\log2\right)\\
\mathcal{N}\left(1\right) & = & \left.-\frac{\partial}{\partial\nicefrac{1}{r}}g\left(r\right)\right|_{r=1}=-\alpha\left(\log\log2-\log K\right)\log2
\end{eqnarray}

From these, we can find the leading terms of the number of different
dominant internal representations:
\begin{equation}
e^{N\mathcal{N}\left(1\right)}=\left(\frac{K}{\log2}\right)^{N\alpha\log2}
\end{equation}
and their volume:
\begin{equation}
e^{-Nw\left(1\right)}=2^{N\left(1-\alpha\right)}\left(\frac{\log2}{K}\right)^{N\alpha\log2}
\end{equation}

\section{Appendix: time discretization\label{sec:Appendix:-time-discretization}}

\subsection{Modified BP-inspired learning scheme for continuous inputs}

The learning protocol presented in section~\ref{sub:BPI} can be easily
generalized to the case in which the input patterns $\xi_{it}^\mu$ are
not binary, but positive and continuous: the only required change is
that the update rules, rather then being applied only to those
synapses for which $\xi_{it^\star}^\mu=1$, are applied to all synapses
with probability $p_i^\mu=\min\left(\xi_{it^\star}^\mu, 1\right)$. Therefore,
the actions taken upon determining $t^\star$ and the value
$\Phi^\mu$ are:
\begin{description}
\item [{$\Phi^{\mu}>1$}]: do nothing
\item [{$0<\Phi^{\mu}\le1$}]: with probability $r$, update synapses for
which $J_{i}=\sigma_{\exp}^{\mu}$, each with probability $p_i^\mu$;
with probability $\left(1-r\right)$ do nothing
\item [{$\Phi^{\mu}\le0$}]: update all synapses, each with probability $p_i^{\mu}$
\end{description}
In order for this generalization to be effective without furher
modifications of the algorithm, it is crucial that a normalization
step is applied to the inputs (see next section).

As an additional generalization, we also introduce a robustness
parameter $\rho$ and re-define
$\Phi^{\mu}=\sigma_{\exp}^{\mu}\Delta_{t^{\star}}^{\mu} - \rho \theta$,
where $\theta$ is the firing threshold: this forces the learning algorithm
to seek solutions in which the depolarization is far from the
threshold. In the numerical experiments described in section~\ref{sub:Continuous}
we used the value $\rho=0.2$, increasing it from $0$ in steps of $0.01$ for $1000$
iterations at each step; the other parameters of the model used in those
tests were $N=1000$, $K=50$, $h_{\max}=25$ and $r=0.3$.

\subsection{Pattern time-discretization}

In this section we describe the time-discretization process mentioned
in section~\ref{sub:Continuous}: we consider a continuous-time model
as described in the Introduction; then, for any given input spike
train, we compute the post-synaptic-potential trace
$R_i^\mu\left(t\right)=\sum_{t_{i}^\mu<t}v\left(t-t_{i}^\mu\right)$,
where $v\left(t\right)$ is the temporal kernel of the membrane. We
divide the time window $T$ in $K$ equal bins, and for each bin $k$ we
compute the input $\xi_{ik}^\mu$ as the fraction of the membrane
kernel in that bin:
\begin{equation}
    \xi_{ik}^\mu = \frac{1}{t_m-t_s} \int_{k-\textrm{th}\ \textrm{bin}}dt\ R_i^\mu\left(t\right)
\end{equation}
where we used the fact that $\int_0^\infty dt\ v\left(t\right)=t_m-t_s$
with out choice of $v$.

Note that the resulting $\xi_{ik}^\mu$ can be greater than $1$, but this is rare
under the sparsity regime which we considered.

The time-discretized patterns can then be passed to the discrete algorithm for
deriving a vector of synaptic weights, which in turn can be tested on the
original model. In our numerical experiments, we generated input spike
trains by a Poisson process with a rate chosen as to obtain
the correct value of the input frequency $f$ (see section~\ref{par:Model})
after the discretization in $K$ time bins. When testing the solution,
we used the value of the firing threshold for the continuous unit which
gave the lowest number of errors.

\bibliographystyle{iopart-num}

\begin{thebibliography}{10}
\expandafter\ifx\csname url\endcsname\relax
  \def\url#1{{\tt #1}}\fi
\expandafter\ifx\csname urlprefix\endcsname\relax\def\urlprefix{URL }\fi
\providecommand{\eprint}[2][]{\url{#2}}

\bibitem{gutig_tempotron:_2006}
G{\"u}tig R and Sompolinsky H 2006 {\em Nature Neuroscience\/} {\bf 9} 420--428
  ISSN 1097-6256
  \urlprefix\url{http://www.nature.com/neuro/journal/v9/n3/abs/nn1643.html}

\bibitem{johansson_first_2004}
Johansson R~S and Birznieks I 2004 {\em Nature Neuroscience\/} {\bf 7} 170--177
  ISSN 1097-6256
  \urlprefix\url{http://www.nature.com/neuro/journal/v7/n2/full/nn1177.html}

\bibitem{decharms_primary_1996}
{deCharms} R~C and Merzenich M~M 1996 {\em Nature\/} {\bf 381} 610--613 ISSN
  0028-0836

\bibitem{meister_concerted_1995}
Meister M, Lagnado L and Baylor D~A 1995 {\em Science\/} {\bf 270} 1207--1210
  ISSN 0036-8075

\bibitem{wehr_odour_1996}
Wehr M and Laurent G 1996 {\em Nature\/} {\bf 384} 162--166 ISSN 0028-0836

\bibitem{rubin_theory_2010}
Rubin R, Monasson R and Sompolinsky H 2010 {\em Physical Review Letters\/} {\bf
  105} 218102
  \urlprefix\url{http://link.aps.org/doi/10.1103/PhysRevLett.105.218102}

\bibitem{braunstein_learning_2006}
Braunstein A and Zecchina R 2006 {\em Physical Review Letters\/} {\bf 96}
  030201 \urlprefix\url{http://prl.aps.org/abstract/PRL/v96/i3/e030201}

\bibitem{baldassi_efficient_2007}
Baldassi C, Braunstein A, Brunel N and Zecchina R 2007 {\em Proceedings of the
  National Academy of Sciences\/} {\bf 104} 11079--11084 ISSN 0027-8424,
  1091-6490 {PMID:} 17581884
  \urlprefix\url{http://www.pnas.org/content/104/26/11079}

\bibitem{baldassi_generalization_2009}
Baldassi C 2009 {\em Journal of Statistical Physics\/} {\bf 136} ISSN 0022-4715
  (Print) 1572-9613 (Online)
  \urlprefix\url{http://www.springerlink.com/content/r07772l167526045/}

\bibitem{mezard_space_1989}
M{\'e}zard, M 1989 {\em Journal of Physics A: Mathematical and General\/} {\bf 22} 2181--2190
  ISSN 0305-4470, 1361-6447

\bibitem{braunstein_encoding_2007}
Braunstein A, Kayhan F, Montorsi G and Zecchina R 2007 Encoding for the
  blackwell channel with reinforced belief propagation {\em {IEEE}
  International Symposium on Information Theory ({ISIT07)}\/} pp 1891--1895

\bibitem{bailly-bechet_finding_2010}
Bailly-Bechet M, Borgs C, Braunstein A, Chayes J, Dagkessamanskaia A, Fran{\c
  c}ois J and Zecchina R 2010 {\em Proceedings of the National Academy of
  Sciences\/} {\bf 108} 882--887 ISSN 0027-8424
  \urlprefix\url{http://www.pnas.org/content/108/2/882.short}

\bibitem{mezard_analytic_2002}
M{\'e}zard M, Parisi G and Zecchina R 2002 {\em Science\/} {\bf 297} 812 -- 815
  ISSN 10959203

\bibitem{braunstein_survey_2005}
Braunstein A, M{\'e}zard M and Zecchina R 2005 {\em Random Structures and
  Algorithms\/} {\bf 27} 201--226

\bibitem{monasson_weight_1996}
Monasson R and Zecchina R 1996 {\em Physical Review Letters\/} {\bf 76}
  2205--2205
  \urlprefix\url{http://link.aps.org/doi/10.1103/PhysRevLett.76.2205.3}

\bibitem{cocco_analytical_1996}
Cocco S, Monasson R and Zecchina R 1996 {\em Physical Review E\/} {\bf 54}
  717--736 \urlprefix\url{http://link.aps.org/doi/10.1103/PhysRevE.54.717}

\end{thebibliography}

\providecommand{\newblock}{}

\end{document}